\documentclass{Interspeech2024}

\usepackage{cite}
\usepackage{tipa}

\interspeechcameraready

\title{Articulatory Configurations across Genders and Periods in French Radio and TV archives}

\name[affiliation={1}]{Benjamin}{Elie}
\name[affiliation={2}]{David}{Doukhan}
\name[affiliation={2,3}]{Rémi}{Uro}
\name[affiliation={3}]{Lucas}{Ondel-Yang}
\name[affiliation={3}]{Albert}{Rilliard}
\name[affiliation={2,3}]{Simon}{Devauchelle}

\address{
  $^1$The University of Edinburgh, UK
  $^2$INA, France
  $^3$Université Paris Saclay, CNRS, LISN}
\email{benjamin.elie@ed.ac.uk, \{ddoukhan,ruro\}@ina.fr, \{lucas.ondel-yang,albert.rilliard,simon.devauchelle\}@lisn.fr}


\begin{document}


\maketitle

\begin{abstract}
This paper studies changes in articulatory configurations across genders and periods using an inversion from acoustic to articulatory parameters. From a diachronic corpus based on French media archives spanning 60 years from 1955 to 2015, automatic transcription and forced alignment allowed extracting the central frame of each vowel. More than one million frames were obtained from over a thousand speakers across gender and age categories. Their formants were used from these vocalic frames to fit the parameters of Maeda's articulatory model. Evaluations of the quality of these processes are provided. We focus here on two parameters of Maeda’s model linked to total vocal tract length: the relative position of the larynx (higher for females) and the lips protrusion (more protruded for males). Implications for voice quality across genders are discussed. The effect across periods seems gender independent; thus, the assertion that females lowered their pitch with time is not supported.
\end{abstract}

\section{Introduction}

Attributing a gender to an interlocutor proves fundamental in many languages and socio-cultural systems to select grammatical characteristics or for social cohesion, e.g., for the complex system of Thai honorifics~\cite{Tawilapakul_2022} or to select adequate terms of address~\cite{RebolloCouto_Lopes_2011}. 
Gender categories are thus rapidly decided when interacting with someone, and voice carries important cues for this attribution~\cite{Childers_Wu_1991, Leung_Oates_Chan_2018, Weirich_Simpson_2018}.
In relation to this, each speaker builds their voice during infancy, before \cite{Guzman_2014, Cartei_Banerjee_Hardouin_Reby_2019} 
and during \cite{Markova_2016, Yamauchi_2015} puberty, and along adulthood  \cite{Ohara_2001}, to fit their vocal identity to their own gender, among other social characteristics.
The construction of voice and gender display is notably linked to cultural and social aspects~\cite{vanBezooijen_1995}. 
A voice also cues other behavioral aspects, such as dominance, which is notably related to lower pitch \cite{Puts_Gaulin_Verdolini_2006}; this has symbolic implications at the linguistic and social levels (see Ohala's "Frequency Code" \cite{Ohala_1994}).

The complex use of vocal characteristics in social interactions imposes strong demands on one's voice to match the expectations for their social position. Such expectations are particularly important and significant in the case of female speakers in public arenas, such as in politics, as acoustic cues deemed "feminine" (typically higher pitch) are associated with features irrelevant to senior and decision-making positions, which are associated with "masculine," lower pitched, voice characteristics~\cite{vanBezooijen_1995, Puts_Gaulin_Verdolini_2006}. 
In the French political arena (but not only), the voice of female politicians (and not the content of their speech) is thus often commented on and criticized for supposedly not being adequate to their duties \cite{Coulomb-Gully_Rennes_2010, Coulomb-Gully_2022}. This led some politicians to follow coaching sessions to deepen their voice, while others feel the need to change their accent to fit different positions \cite{Riverin-Coutlée_Harrington_2022}. In another domain where voices play a fundamental role, cinema and TV actresses receive varying demands for their voices, which may sometimes be contradictory and depend on the context (dubbing having different requirements than an original production) and typically on the individual representations of "femininity" \cite{LeFevre-Berthelot_2015}. 

In this landscape, some claim that female voices have lowered over time. 
These claims are sourced from a study comparing two groups of female Australian English speakers in their 20s over 50 years  \cite{Pemberton_McCormack_Russell_1998}. 
A more recent study cites a possible lowering of female voices, but without diachronic data \cite{Berg_Fuchs_Wirkner_Loeffler_Engel_Berger_2017}. 
These trends are debated, with an "inconsequential" trend over time but a clearer difference across styles found in \cite{Hollien_Hollien_DeJong_1997}. 
Few studies focus on French \cite{Suire_Barkat-Defradas_2020}, with unclear results, and not controlling for the speakers' age, for example.

To better understand potential changes in voice across gender and to document the characteristics of female and male voices displayed in the media, we analyze a diachronic, cross-sectional corpus of French media archives spanning over 60 years. 
This work focuses on French national media as a prescriptive source for the language spoken in France and for voice representation for the general population; this leaves out of the scope of this work the voices of the French society under-represented in mainstream medias. 
Media archives are nonetheless interesting, notably because they present excerpts of spontaneous spoken interactions, which have a social impact due to their exposition to a wide audience.
Gender is approached here as a binary category opposing female and male cisgender individuals -- which compose the vast majority of appearances in TV and radio media (with a strong bias toward ethnically homogeneous cisgender male guests \cite{Doukhan_Poels_Rezgui_Carrive_2018}; see, e.g., \cite{Johnson_2006, Hope_Lilley_2022} for non-binary approaches of gender and voice). Our corpus makes an effort to lessen gender bias. To study the pitch of voices in the corpus, we focus here on the averaged articulatory configuration of speakers, estimated through an inversion technique that estimates the parameters of Maeda's articulatory model \cite{maeda1990compensatory} through an inversion process from vowels' formants over large series of vowels. 

For an analysis of pitch through fundamental frequency on the same corpus, see 
\cite{Rilliard_Doukhan_Uro_Devauchelle_2023}.
Inversion techniques, though complex, may lead to interesting trends and help understand the significance of changes in resonances in terms of articulation and gender-specific strategies for voice placement.
The article first presents the dataset and its processing to reliably estimate formants from unscripted media archives. It then dives into the inversion process to estimate Maeda's model parameters, briefly explaining their role and selecting a subset pertinent for voice pitch \cite{Ohala_1994}. Results are then presented and discussed for their relevance to gender and voice in a diachronic perspective.

\section{Methods}
\subsection{Dataset}
The study is based on a corpus of TV and radio samples obtained from the French National Institute of Audiovisual (INA). This corpus contains speech samples of 1025 speakers broadcast during four time periods: 1955-56, 1975-76, 1995-96, and 2015-16. 
Speakers are spread across genders and age categories (20-35, 36-50, 51-65, and over 65 y.o.) according to Table~\ref{tab:speakers}. 
An initial target of 30 speakers for each of the 32 period/gender/age categories was set; it was not met for some categories, especially in the oldest periods, for females, and for the oldest age groups, due to a particularly important under-representation of these demographic categories in the media in the 20th century (that continues today).
Selected speech samples present a high acoustic quality for media archives, mostly free of background noises and music.
This was made possible by applying the selection method proposed by \cite{Uro_2022}, which uses the source separation algorithms proposed by \cite{Hennequin_Khlif_Voituret_Moussallam_2020} to evaluate the relative speech-to-noise ratio, to localize and discard the most noisy excerpts (noise being either music or other noises). 
A target of at least 3 minutes of speech by speaker was set.
The complete corpus represents about 111 hours of speech without pauses.
A complete description of the corpus construction can be found in 
\cite{Uro_2022,Rilliard_Doukhan_Uro_Devauchelle_2023}.

\begin{table}[th]
  \caption{Distribution of the 1025 speakers in categories of Period and Gender (Female / Male) for each age group (lines).}
  \label{tab:speakers}
  \centering
  \begin{tabular}{ r c c c c c c c c }
    \toprule
  Period & \multicolumn{2}{c}{\textbf{1955-56}} & 
        \multicolumn{2}{c}{\textbf{1975-76}} & 
        \multicolumn{2}{c}{\textbf{1995-96}} & 
        \multicolumn{2}{c}{\textbf{2015-16}} \\
    \midrule
  Gender &  F &  M &  F &  M &  F &  M &  F &  M \\
    \midrule
  20-35 & 18 & 41 & 18 & 18 & 30 & 29 & 31 & 31 \\
  36-50 & 21 & 72 & 23 & 42 & 33 & 45 & 30 & 53 \\
  51-65 & 18 & 51 & 27 & 37 & 29 & 48 & 29 & 49 \\
  $>65$ & 19 & 15 & 18 & 25 & 29 & 34 & 31 & 31 \\
    \bottomrule
  \end{tabular}  
\end{table}

To identify vocalic segments, the corpus was automatically transcribed using Whisper~\cite{whisper}. 
The quality of the transcription was estimated on a subset of 81 segments (about 20 minutes) randomly selected from all periods, gender, and age categories.
Four people have manually transcribed the segments without overlap. 
An evaluation of the word error rate (WER) and of the phone error rate (PER) was done (Table~\ref{tab:eval}). Whisper reaches a $12.8\%$ WER on the manually transcribed subset. 

\begin{table}[th]
    \caption{Results of the WER and PER (hits, substitutions, deletions, insertions), obtained on the subset after applying the text normalization. Details of the PER for vowels only are also specified.}
    \label{tab:eval}
    \centering
    \begin{tabular}{lcccc||c}
    \hline
    Edit types   & \textbf{H} & \textbf{S} & \textbf{D} & \textbf{I} & Error rates (\%) \\ \hline \hline
    Words         & 4124 & 197   & 316  & 82 & 12.8 \\ \hline
    Phones & 13828 & 169 & 801 & 205 & 7.9\\
    Oral vowels & 5186 & 89 & 386 &  79 & 9.8\\ \hline
    \end{tabular}
\end{table}

The PER was obtained by running the Montreal Forced Aligner (MFA)\cite{mcauliffe17_interspeech} with its French dictionary and aligner on Whisper's predictions and on the ground truth, dropping to $7.9\%$. The 12 oral vowels of French were selected for the last PER in Table \ref{tab:eval}.
$5.4 \%$ of the vowels in the evaluation subset were removed in the Whisper transcription. This shows a reduced effect of Whisper smoothing its transcripts. 
The vowels most frequently deleted are /\textipa{\oe}/ (46.7\% deletion), /\textipa{\o}/ (44.6\%), /\textipa{\textschwa}/ (8.6\%) and /\textipa{e}/ (7.2\%). 25\% of the deleted phones are derived from the French word for hesitation, ``euh" (pronounced /\textipa{\oe}/ or /\textipa{\o}/) followed by the function words ``et" (6.5\%) and ``de" (4.7\%). 






This leads to 1,139,818 segmented vowels once those unvoiced or longer than 200ms were removed (they represented about 20\% of the complete vowel set).
The central 25ms frame of each vowel was then targeted for the acoustical analysis.

\subsection{Acoustic feature extraction}

The first four formants ($F_1,F_2,F_3,F_4$) were estimated thanks to Praat \cite{Boersma_Weenink_2023} Burg algorithm.
To estimate formants, the method proposed in \cite{Escudero_Boersma_Rauber_Bion_2009} was applied, which consists of a systematic variation of the frequency ceiling by speaker and phone to find an optimal ceiling -- that minimizes the formant variation within each category.
We estimated each speaker's Vocal Tract Length ($VTL$) from the formant obtained on each vowel by applying the equations 1 and 4 from \cite{lammert2015short} (cf. Eqs.~\ref{x:vtl1}\&~\ref{x:vtl2}); the mean of $VTL$ estimations for all vowels of a speaker was kept as the speaker's estimated $VTL$.
\begin{equation}
\phi = \frac{0.089 F_1}{1} + \frac{0.102 F_2}{3} + \frac{0.121 F_3}{5} + \frac{0.669 F_4}{7}
\label{x:vtl1}
\end{equation}
\begin{equation}
VTL = \frac{c}{4 * \phi}
\label{x:vtl2}
\end{equation}
where $F_1,F_2,F_3,F_4$ are the first four formant values, and $c = 34000 cm/s$ is the speed of sound.
These procedures were applied to the raw recordings of the corpus, and then only the values from the central frame of each vowel were kept.


\subsection{Acoustic-to-articulatory inversion}\label{ac2art:sec}

\begin{figure}
    \centering
    \includegraphics[width=0.4\linewidth]{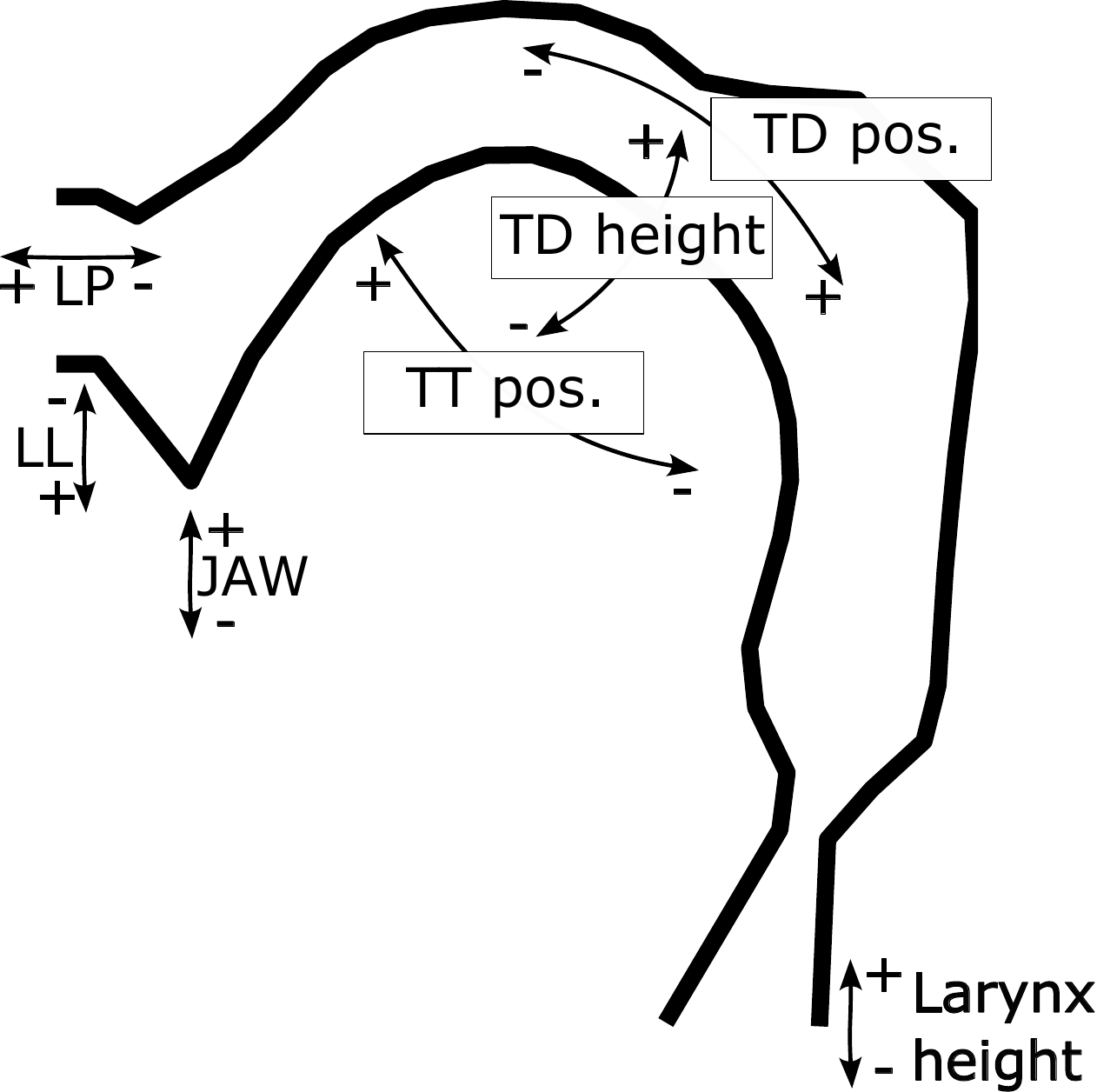}
    \caption{Parameters of the Maeda model with directions of movement towards positive '+' and negative values '-'.}
    \label{maeda_model:fig}
\end{figure}

We performed acoustic-to-articulatory inversion to estimate the articulatory parameters of speakers. The articulatory parameters are the seven parameters of Maeda's articulatory model~\cite{maeda1990compensatory}. The Maeda model generates midsagittal shapes of the vocal tract using seven independent articulatory parameters corresponding to the principal components that explain most of the observed variance in articulatory data. These parameters are expressed in standard deviation around their mean value. These seven parameters are associated with the position of the jaw (JAW), the anteroposterior position and height of the tongue dorsum (TD position and TD height), the vertical position of the tongue tip (TT position), the height of the lower lip (LL), the lip protrusion (LP), and the larynx height (LH), as shown in Figure~\ref{maeda_model:fig}. 

The inversion process was based on formant values: the selected inverted articulatory parameters minimized the difference between generated and observed formant frequencies. For each vowel realization of each speaker, we solved the following optimization problem:
\begin{equation}
\mathbf{x} = \underset{\hat{\mathbf{x}}}{\operatorname{argmin}} \vert \vert \mathbf{f}_\mathrm{obs} - \mathbf{f}_\mathrm{gen}(\hat{\mathbf{x}}) \vert \vert^2,
\label{x:eq}
\end{equation}
where $\mathbf{f}_\mathrm{obs}$ contains the first four observed formants (\textit{i.e.} produced by the speakers, and estimated from the central frame of each vowel), and $\mathbf{f}_\mathrm{gen}(\hat{\mathbf{x}})$ contains the first four formants computed from the midsagittal shapes of the vocal tract generated by the vector $\hat{\mathbf{x}}$ of Maeda's parameters. 


The problem of articulatory inversion is a well-known ill-posed problem: an infinity of articulatory configurations can generate the same formant frequency pattern~\cite{atal1978inversion}. As such, the solution of Eq.~\eqref{x:eq} is not unique. Some authors have included temporal constraints to regularize the problem, e.g., smooth formant trajectories~\cite{Ouni_JASA, Pancha_JASA}, a solution unavailable because we are trying to invert vowels individually. Instead, we kept the solutions that minimize the dispersion in the articulatory space for realizations of the same vowel of a given speaker. 

For that purpose, we proceeded as follows. The Maeda model was first adapted to the speaker's average $VTL$ (estimated as described above) by modifying the model's size correction factor such that the length of the vocal tract of the neutral configuration (\textit{i.e.} the null vector) is that of the speaker. 
Then, for each of the $N$ realizations of a vowel by a speaker, we ran 20 optimizations using the Nelder-Mead algorithm~\cite{nelder1965simplex} initialized at random, resulting in a set of $20N$ solution vectors $\mathbf{x}$. We then computed the weighted mean solution vector $\mathbf{\bar{x}}$ of these solutions, where the weights were computed as the $\log$ of the final cost function, as follows:
\begin{equation}
\mathbf{\bar{x}} = \frac{\sum_{i=1}^{20N} - \log(\vert \vert \mathbf{f}_\mathrm{obs}^{(i)} - \mathbf{f}_\mathrm{gen}({\mathbf{x}_i}) \vert \vert) {\mathbf{x}_i}}{\sum_{i=1}^{20N} - \log(\vert \vert \mathbf{f}_\mathrm{obs}^{(i)} - \mathbf{f}_\mathrm{gen}({\mathbf{x}_i}) \vert \vert)}.
\label{sol:eq}
\end{equation}

Then, for each realization $n$ of the $N$ vowels, among the set of 20 solutions corresponding to $n$, we chose the closest solution to the mean solution $\mathbf{\bar{x}}$ in the articulatory space, namely with the lowest Euclidean distance.

We concentrated here only on two parameters, the larynx height LH and the lip protrusion LP, because they directly affect the overall length of the vocal tract, and thus its resonances. They both are used with known effect in animal behaviors to signal size and related behavioral tendencies -- for example submissive/aggressive behavior respectively linked with smile or protrusion (see the comment on the acoustic origin of smile in \cite{Ohala_1994}), or the lowered larynx of deer used for body size exaggeration during the mating season \cite{TecumsehFitch_Reby_2001}.
These parameters, estimated from an inversion procedure to fit Maeda's model, indicate required changes at both ends of the vocal tract to fit the observed formants, and not directly the vocal tract length but only a modification of the average size.



\subsection{Statistical analysis process}

%
%
Based on these two predicted articulatory parameters, LH and LP, used as dependent variables, we fit linear mixed models using R's \textit{lme4} library \cite{Bates_2015}, with the following fixed factors: the speakers' Age (
note that the four age categories used to select the speakers are not used at the model fitting stage), the speakers' Gender (Female/Male) and the time Period (1955-56, 1975-76, 1995-96, 2015-16). The 1025 Speakers and the 12 oral Vowels were treated as nested random effects (with vowels nested within each specific speaker).
All double and triple interactions between the fixed factors were included in the model.
A model simplification procedure was then applied to each model, using the \textit{step()} function \cite{lmertest}.
The minimal adequate models after these simplification steps are described in Eq.~\ref{mlh:eq}, where $DV$ is the dependent variable, either LH or the LP, explained by Age, Gender (and their interaction) plus Period, and the two random factors Vowel nested in Speaker.
\begin{equation}
DV \sim Age \ast Gender + Period + (1 | Speaker/Vowel) 
\label{mlh:eq}
\end{equation}

Table~\ref{tab:anova} details the ANOVA table for both models. The random factors had an important role but for the Speaker factor for the larynx height.

\begin{table}[th]
    \caption{Analysis of Deviance Table (Type II Wald chisquare tests) for the models based on Larynx Height (top) and Lip protrusion (bottom), reporting for each factor and the interaction their estimated $\chi^2$, degree of freedom and associated probability.}
    \label{tab:anova}
    \centering
    \begin{tabular}{lrcl}
    \hline
Factor  &  Chisq  & Df & Pr(>Chisq)  \\   
\hline
Age     &   4.16  & 1  &   $< 0.05$  \\ 
Gender  & 455.59  & 1  &   $< 0.001$ \\
Period  &  57.06  & 3  &   $< 0.001$ \\
Age:Gen &  21.28  & 1  &   $< 0.001$ \\
\hline
\hline
Factor  &  Chisq  & Df & Pr(>Chisq)     \\
\hline
Age     &     0.03 & 1 &   $  0.852 $ \\      
Gender  &   145.82 & 1 &   $< 0.001$ \\ 
Period  &    54.09 & 3 &   $< 0.001$ \\ 
Age:Gen &     4.43 & 1 &   $< 0.05$ \\  
\hline
    \end{tabular}
\end{table}

\section{Results}\label{results:sec}


\subsection{Model of larynx height}

\begin{figure}[ht!]
\centering
\includegraphics[width=\linewidth]{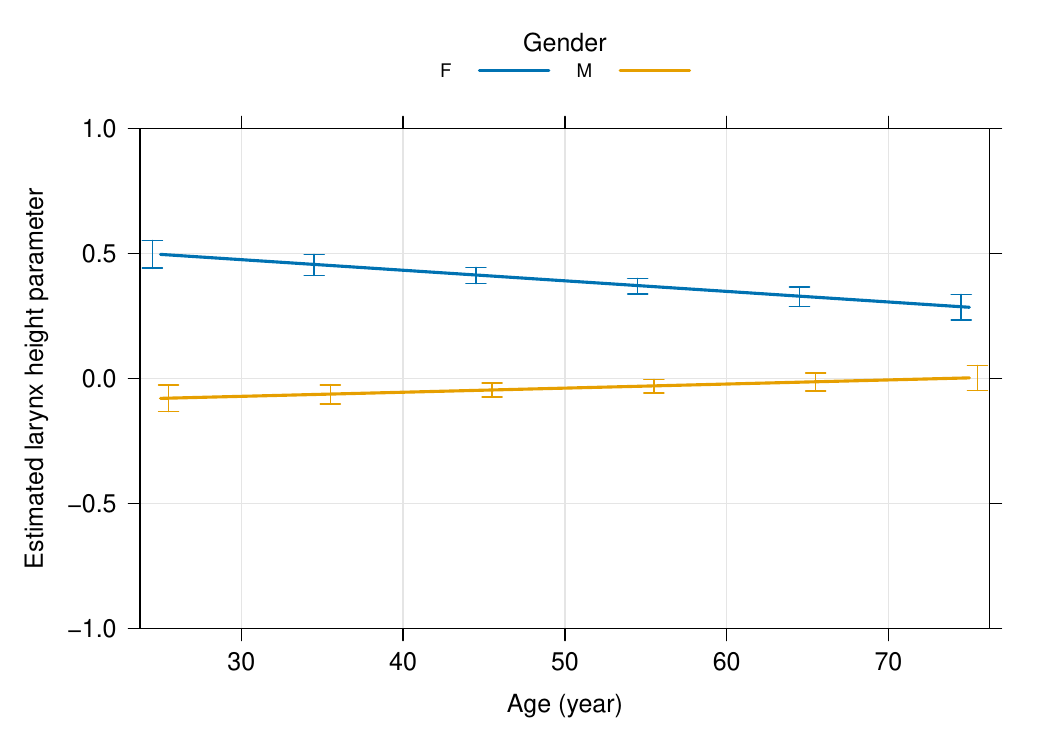}
\includegraphics[width=\linewidth]{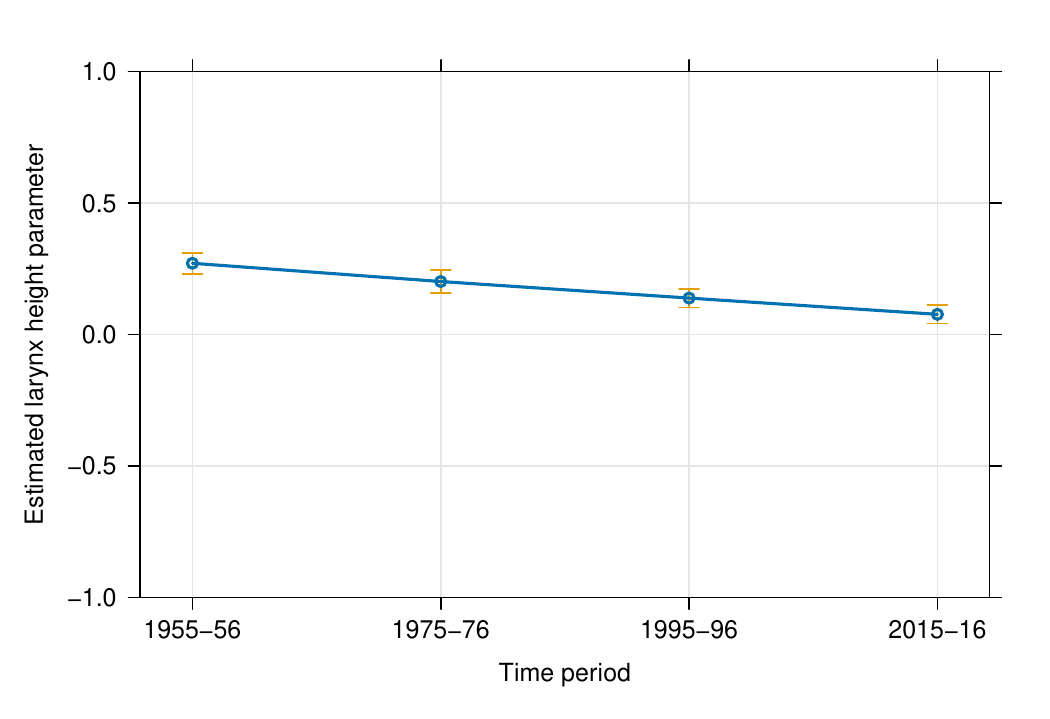}
\caption{Effects on the LH parameter of (top) speaker Age (x-axis) and Gender (curves), and (bottom) period (x-axis).}
\label{LHAG:fig}
\end{figure}

Let's recall the Maeda's model parameters of interest LH and LP are in arbitrary unit (result of a PCA output), and varying between -3 and +3.
The model fit on the estimation of larynx height showed a significant change over age, dependent on the speaker's age (see Fig.~\ref{LHAG:fig}, top panel): the effect is reverse for each gender, with a decreasing value for females with age (i.e., the larynx tends to be lower with older age), while LH rises with age for males. 
The main difference observed on the LH parameter is related to Gender, with females tending to raise their larynx (mean value for $LH = 0.37$), while males slightly tend to lower it ($LH = -0.03$).
The mean LH position also changed across the time period ($LP$ decreases as depicted in the bottom panel of  Fig.~\ref{LHAG:fig}), with lower predicted LH values from the 1955's until 2015. Importantly, this change is not dependent on gender and thus applies to both female and male speakers. The lowering of the larynx across the period is about -0.07 cm for each period, which corresponds to a lengthening of less than 0.5\% on the vocal tract for 60 years -- or a really small amount of change (considering that the maximal change possible with the model amount to more than 10\% of the total tract length).

\subsection{Model of lip protrusion}
\begin{figure}[ht!]
\centering
\includegraphics[width=\linewidth]{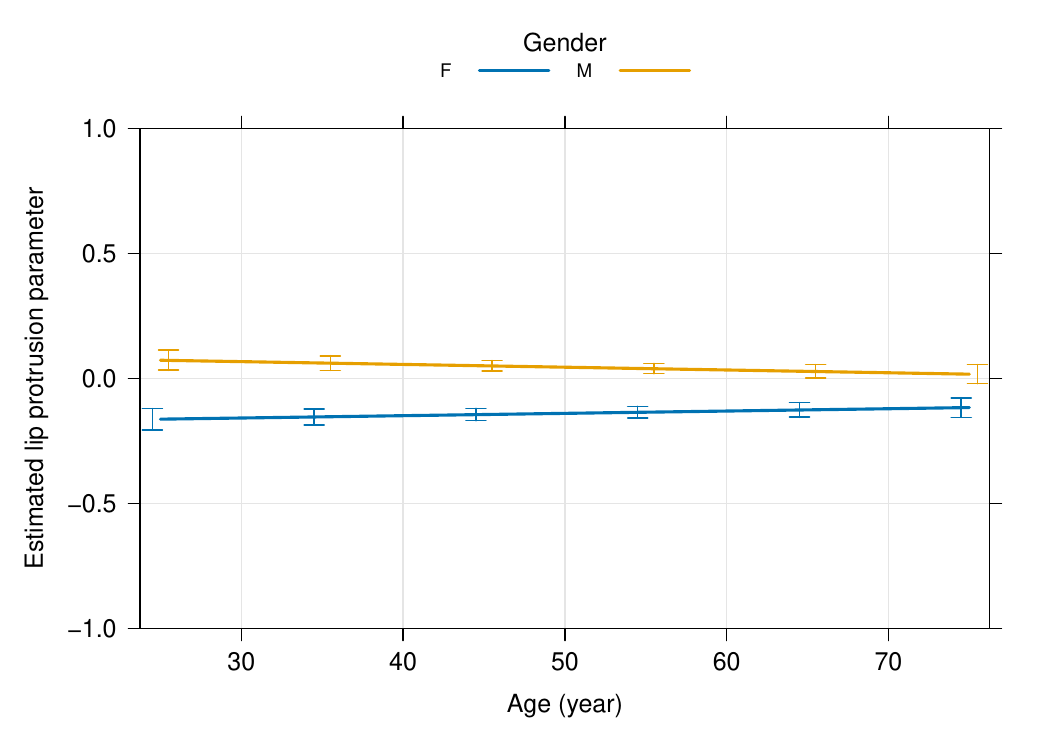}
\caption{Effects of speaker Age (x-axis) and Gender (curves) on the LP parameter.}
\label{LPAG:fig}
\end{figure}

The values fitted for LP also change according to the speaker Age, across gender (Fig.~\ref{LPAG:fig}. But for this parameter, males tend to produce more protrusion ($LP = 0.04$) while females tend to have more stretched lips ($LP = 0.14$). The change with Age is also opposed across Genders and tends to neutralize gender differences.
Lip protrusion also differs across periods (without link to gender), with a tendency to more protrusions in the two more recent periods ($LP \simeq 0$), compared to 1955 and 1975 ($LP \simeq -0.1$).
Here also, the size of these changes is very limited, with the model also predicting changes of less than 1\% of the total vocal tract length.

\section{Discussion and Conclusions}

We applied an acoustic inversion technique to estimate articulatory parameters from a large corpus of media speech excerpts. 
The methods, based on the automatic processing of large quantities of spontaneous speech, proved efficient: the automatic transcription tools do a robust job of extracting most of the speech, and the presented evaluation showed that, word deletions by Whisper applies mostly to hesitations and repetitions. 
The quality in terms of phonetic alignment is better than at the lexical level. 

Regarding the question raised for this dataset -- do females in French media have vocal characteristics that have changed over this time period? -- the answer is contrasted.
We observed changes in terms of adjustment to the vocal tract, and the most important differences reported here are related to gender, with females tending to raise their larynx and stretch their lips, while males tend to lower their larynx and protrude their lips (note that we are talking of the predictions of an inversion model, that reflect at best modest articulatory changes, one shall not exaggerate these articulatory differences).
Such gender-specific differences are reminiscent of the strategies described by Ohala in relation to the Frequency Code \cite{Ohala_1994}. 
However, these gender differences do not change over time periods, but they change with the speakers' age: older speakers tend to show fewer gender-specific differences.
Over periods, the evolution is not gender-specific but exists and goes towards lower resonances (lower larynx, more protrusion).

So have females lowered resonances over periods? Yes, but males also.
The analysis of this corpus does not support the fact that females only have lowered their voices over time -- at least for the population represented in this corpus (composed of individuals invited in media and referenced in the INA archives, so who reached some fame: almost all of them have a Wikipedia page, for example).
If a change over periods occurred, it applies to both genders, and it may be interpreted as a change in the situational setting where speakers were recorded, typically with a microphone farther away from the mouth in older periods.
The distance to the microphone relate to the use of a more projected voice, requesting a higher vocal effort~\cite{Lienard_2019}, that tends to raise the fundamental frequency and the first formant \cite{Titze_Sundberg_1992,Rilliard_2018}. 
A closer microphone and softer voices may explain the lower pitch in a more recent broadcast. This does not constitute a change in voice for a category of the population, but rather a stylistic change, as already noted for journalists by, e.g., \cite{Boula_2012}. 
The more important presence of older females (and males) over 60 years of age in more recent periods (a fact confirmed during the documentation work made to create this corpus) may also be part of the explanation, as female voices deepen with age.
Finally, it is also possible that the pressure for a lower voice is felt by certain categories of the female population (typically those having political or economic responsibilities), and a split of the speaker across their occupation may be worth further investigation.

\section{Acknowledgements}
This work has been partially funded by the French National Research Agency (GEM project - ANR-19-CE38-0012).

\bibliographystyle{IEEEtran}
\bibliography{mybib}

\begin{thebibliography}{10}
\providecommand{\url}[1]{#1}
\csname url@samestyle\endcsname
\providecommand{\newblock}{\relax}
\providecommand{\bibinfo}[2]{#2}
\providecommand{\BIBentrySTDinterwordspacing}{\spaceskip=0pt\relax}
\providecommand{\BIBentryALTinterwordstretchfactor}{4}
\providecommand{\BIBentryALTinterwordspacing}{\spaceskip=\fontdimen2\font plus
\BIBentryALTinterwordstretchfactor\fontdimen3\font minus \fontdimen4\font\relax}
\providecommand{\BIBforeignlanguage}[2]{{%
\expandafter\ifx\csname l@#1\endcsname\relax
\typeout{** WARNING: IEEEtran.bst: No hyphenation pattern has been}%
\typeout{** loaded for the language `#1'. Using the pattern for}%
\typeout{** the default language instead.}%
\else
\language=\csname l@#1\endcsname
\fi
#2}}
\providecommand{\BIBdecl}{\relax}
\BIBdecl

\bibitem{Tawilapakul_2022}
U.~Tawilapakul, ``\BIBforeignlanguage{en}{{Face threatening and speaker presuppositions: The case of feminine polite particles in Thai}},'' \emph{\BIBforeignlanguage{en}{Journal of Pragmatics}}, vol. 195, p. 69–90, 2022.

\bibitem{RebolloCouto_Lopes_2011}
L.~Rebollo~Couto and C.~R. Lopes, Eds., \emph{\BIBforeignlanguage{por spa}{As formas de tratamento em português e em espanhol: variação, mudança e funções conversacionais}}.\hskip 1em plus 0.5em minus 0.4em\relax Niterói, RJ: Editora da UFF, 2011.

\bibitem{Childers_Wu_1991}
D.~G. Childers and K.~Wu, ``\BIBforeignlanguage{en}{{Gender recognition from speech. Part II: Fine analysis}},'' \emph{\BIBforeignlanguage{en}{J. Acoust. Soc. America}}, vol.~90, no.~4, p. 1841–1856, 1991.

\bibitem{Leung_Oates_Chan_2018}
Y.~Leung, J.~Oates, and S.~P. Chan, ``\BIBforeignlanguage{en}{Voice, articulation, and prosody contribute to listener perceptions of speaker gender: {A} systematic review and meta-analysis},'' \emph{\BIBforeignlanguage{en}{Journal of Speech, Language, and Hearing Research}}, vol.~61, no.~2, p. 266–297, 2018.

\bibitem{Weirich_Simpson_2018}
M.~Weirich and A.~P. Simpson, ``\BIBforeignlanguage{en}{Gender identity is indexed and perceived in speech},'' \emph{\BIBforeignlanguage{en}{PLOS ONE}}, vol.~13, no.~12, 2018.

\bibitem{Guzman_2014}
M.~Guzman, D.~Muñoz, M.~Vivero, N.~Marín, M.~Ramírez, M.~T. Rivera, C.~Vidal, J.~Gerhard, and C.~González, ``\BIBforeignlanguage{en}{Acoustic markers to differentiate gender in prepubescent children’s speaking and singing voice},'' \emph{\BIBforeignlanguage{en}{International Journal of Pediatric Otorhinolaryngology}}, vol.~78, no.~10, p. 1592–1598, 2014.

\bibitem{Cartei_Banerjee_Hardouin_Reby_2019}
V.~Cartei, R.~Banerjee, L.~Hardouin, and D.~Reby, ``\BIBforeignlanguage{en}{The role of sex‐related voice variation in children’s gender‐role stereotype attributions},'' \emph{\BIBforeignlanguage{en}{British Journal of Developmental Psychology}}, vol.~37, no.~3, p. 396–409, 2019.

\bibitem{Markova_2016}
D.~Markova, L.~Richer, M.~Pangelinan, D.~H. Schwartz, G.~Leonard, M.~Perron, G.~Pike, S.~Veillette, M.~M. Chakravarty, Z.~Pausova, and T.~Paus, ``\BIBforeignlanguage{en}{Age- and sex-related variations in vocal-tract morphology and voice acoustics during adolescence},'' \emph{\BIBforeignlanguage{en}{Hormones and Behavior}}, vol.~81, p. 84–96, 2016.

\bibitem{Yamauchi_2015}
A.~Yamauchi, H.~Yokonishi, H.~Imagawa, K.-I. Sakakibara, T.~Nito, N.~Tayama, and T.~Yamasoba, ``\BIBforeignlanguage{en}{Quantitative analysis of digital videokymography: {A} preliminary study on age- and gender-related difference of vocal fold vibration in normal speakers},'' \emph{\BIBforeignlanguage{en}{Journal of Voice}}, vol.~29, no.~1, p. 109–119, 2015.

\bibitem{Ohara_2001}
Y.~Ohara, \emph{{Finding one’s voice in Japanese: A study of the pitch levels of L2 users}}.\hskip 1em plus 0.5em minus 0.4em\relax Berlin: de Gruyter, 2001.

\bibitem{vanBezooijen_1995}
R.~van Bezooijen, ``\BIBforeignlanguage{en}{Sociocultural aspects of pitch differences between {Japanese and Dutch} women},'' \emph{\BIBforeignlanguage{en}{Language and Speech}}, vol.~38, no.~3, p. 253–265, 1995.

\bibitem{Puts_Gaulin_Verdolini_2006}
D.~A. Puts, S.~J. Gaulin, and K.~Verdolini, ``\BIBforeignlanguage{en}{Dominance and the evolution of sexual dimorphism in human voice pitch},'' \emph{\BIBforeignlanguage{en}{Evolution and Human Behavior}}, vol.~27, no.~4, p. 283–296, 2006.

\bibitem{Ohala_1994}
J.~J. Ohala, \emph{The frequency code underlies the sound-symbolic use of voice pitch}.\hskip 1em plus 0.5em minus 0.4em\relax Cambridge University Press, 1994, p. 325–347.

\bibitem{Coulomb-Gully_Rennes_2010}
M.~Coulomb-Gully and J.~Rennes, ``\BIBforeignlanguage{fr}{{Genre, politique et analyse du discours. Une tradition épistémologique française gender blind}},'' \emph{\BIBforeignlanguage{fr}{Mots}}, no.~94, p. 175–182, 2010.

\bibitem{Coulomb-Gully_2022}
M.~Coulomb-Gully, \emph{Sexisme sur la voix publique: femmes, éloquence et politique}.\hskip 1em plus 0.5em minus 0.4em\relax La Tour-d'Aigues: Éd. de l’Aube, 2022.

\bibitem{Riverin-Coutlée_Harrington_2022}
J.~Riverin-Coutlée and J.~Harrington, ``\BIBforeignlanguage{en}{{Phonetic change over the career: A case study}},'' \emph{\BIBforeignlanguage{en}{Linguistics Vanguard}}, vol.~8, no.~1, 2022.

\bibitem{LeFevre-Berthelot_2015}
A.~Le~Fèvre-Berthelot, ``Doublage au féminin~: transposer le genre,'' \emph{Genre en séries}, no.~2, p. 50–72, Jun. 2015.

\bibitem{Pemberton_McCormack_Russell_1998}
C.~Pemberton, P.~McCormack, and A.~Russell, ``\BIBforeignlanguage{en}{{Have women’s voices lowered across time? A cross sectional study of Australian women’s voices}},'' \emph{\BIBforeignlanguage{en}{Journal of Voice}}, vol.~12, no.~2, 1998.

\bibitem{Berg_Fuchs_Wirkner_Loeffler_Engel_Berger_2017}
M.~Berg, M.~Fuchs, K.~Wirkner, M.~Loeffler, C.~Engel, and T.~Berger, ``\BIBforeignlanguage{en}{The speaking voice in the general population: {N}ormative data and associations to sociodemographic and lifestyle factors},'' \emph{\BIBforeignlanguage{en}{J. of Voice}}, vol.~31, no.~2, pp. 13--24, 2017.

\bibitem{Hollien_Hollien_DeJong_1997}
H.~Hollien, P.~A. Hollien, and G.~De~Jong, ``\BIBforeignlanguage{en}{Effects of three parameters on speaking fundamental frequency},'' \emph{\BIBforeignlanguage{en}{J. Acoust. Soc. America}}, vol. 102, no.~5, p. 2984–2992, Nov. 1997.

\bibitem{Suire_Barkat-Defradas_2020}
A.~Suire and M.~Barkat-Defradas, ``{Evolution of human pitch: Preliminary analyses in the French population using INA audiovisual archives of Vox Pops},'' in \emph{IASA-FIAT/IFTA Conference}, 2020.

\bibitem{Doukhan_Poels_Rezgui_Carrive_2018}
D.~Doukhan, G.~Poels, Z.~Rezgui, and J.~Carrive, ``\BIBforeignlanguage{en}{Describing gender equality in french audiovisual streams with a deep learning approach},'' \emph{\BIBforeignlanguage{en}{VIEW Journal of European Television History and Culture}}, vol.~7, no.~14, p. 103, 2018.

\bibitem{Johnson_2006}
K.~Johnson, ``\BIBforeignlanguage{en}{{Resonance in an exemplar-based lexicon: The emergence of social identity and phonology}},'' \emph{\BIBforeignlanguage{en}{Journal of Phonetics}}, vol.~34, no.~4, p. 485–499, 2006.

\bibitem{Hope_Lilley_2022}
M.~Hope and J.~Lilley, ``\BIBforeignlanguage{en}{Gender expansive listeners utilize a non-binary, multidimensional conception of gender to inform voice gender perception},'' \emph{\BIBforeignlanguage{en}{Brain and Language}}, vol. 224, 2022.

\bibitem{maeda1990compensatory}
S.~Maeda, ``{Compensatory articulation during speech: Evidence from the analysis and synthesis of vocal-tract shapes using an articulatory model},'' in \emph{Speech production and speech modelling}.\hskip 1em plus 0.5em minus 0.4em\relax Springer, 1990, pp. 131--149.

\bibitem{Rilliard_Doukhan_Uro_Devauchelle_2023}
A.~Rilliard, D.~Doukhan, R.~Uro, and S.~Devauchelle, ``{Evolution of voices in French audiovisual media across genders and age in a diachronic perspective},'' in \emph{Proceedings of the 20th ICPhS}, 2023.

\bibitem{Uro_2022}
R.~Uro, D.~Doukhan, A.~Rilliard, L.~Larcher, A.-C. Adgharouamane, M.~Tahon, and A.~Laurent, ``A semi-automatic approach to create large gender- and age-balanced speaker corpora: {U}sefulness of speaker diarization \& identification.'' in \emph{LREC'22}, 2022.

\bibitem{Hennequin_Khlif_Voituret_Moussallam_2020}
R.~Hennequin, A.~Khlif, F.~Voituret, and M.~Moussallam, ``{Spleeter: A fast and efficient music source separation tool with pre-trained models},'' \emph{Journal of Open Source Software}, vol.~5, no.~50, p. 2154, 2020.

\bibitem{whisper}
A.~Radford, J.~W. Kim, T.~Xu, G.~Brockman, C.~McLeavey, and I.~Sutskever, ``Robust speech recognition via large-scale weak supervision,'' 2022.

\bibitem{mcauliffe17_interspeech}
M.~McAuliffe, M.~Socolof, S.~Mihuc, M.~Wagner, and M.~Sonderegger, ``{Montreal Forced Aligner: Trainable Text-Speech Alignment Using Kaldi},'' in \emph{Interspeech}, 2017, pp. 498--502.

\bibitem{Boersma_Weenink_2023}
P.~Boersma and D.~Weenink, ``{Praat: Doing phonetics by computer [Computer program]. Version 6.3.20},'' 2023.

\bibitem{Escudero_Boersma_Rauber_Bion_2009}
P.~Escudero, P.~Boersma, A.~S. Rauber, and R.~A.~H. Bion, ``\BIBforeignlanguage{en}{{A cross-dialect acoustic description of vowels: Brazilian and European Portuguese}},'' \emph{\BIBforeignlanguage{en}{J. Acoust. Soc. America}}, vol. 126, no.~3, p. 1379–1393, Sep. 2009.

\bibitem{lammert2015short}
A.~C. Lammert and S.~S. Narayanan, ``On short-time estimation of vocal tract length from formant frequencies,'' \emph{PloS one}, vol.~10, no.~7, 2015.

\bibitem{atal1978inversion}
B.~S. Atal, J.~J. Chang, M.~V. Mathews, and J.~W. Tukey, ``Inversion of articulatory-to-acoustic transformation in the vocal tract by a computer-sorting technique,'' \emph{J. Acoust. Soc. America}, vol.~63, no.~5, pp. 1535--1555, 1978.

\bibitem{Ouni_JASA}
S.~Ouni and Y.~Laprie, ``Modeling the articulatory space using a hypercube codebook for acoustic-to-articulatory inversion,'' \emph{J. Acoust. Soc. America}, vol. 118(1), pp. 444--460, 2005.

\bibitem{Pancha_JASA}
S.~Panchapagesan and A.~Alwan, ``A study of acoustic-to-articulatory inversion of speech by analysis-by-synthesis using chain matrices and the {M}aeda articulatory model,'' \emph{J. Acoust. Soc. America}, vol. 129(4), pp. 2144--2162, 2011.

\bibitem{nelder1965simplex}
J.~A. Nelder and R.~Mead, ``A simplex method for function minimization,'' \emph{The computer journal}, vol.~7, no.~4, pp. 308--313, 1965.

\bibitem{TecumsehFitch_Reby_2001}
W.~Tecumseh~Fitch and D.~Reby, ``\BIBforeignlanguage{en}{The descended larynx is not uniquely human},'' \emph{\BIBforeignlanguage{en}{Proc. of the Royal Society of London. Series B: Biological Sciences}}, vol. 268, no. 1477, p. 1669–1675, 2001.

\bibitem{Bates_2015}
D.~Bates, M.~Mächler, B.~Bolker, and S.~Walker, ``\BIBforeignlanguage{en}{Fitting linear mixed-effects models using lme4},'' \emph{\BIBforeignlanguage{en}{Journal of Statistical Software}}, vol.~67, no.~1, 2015.

\bibitem{lmertest}
A.~Kuznetsova, P.~B. Brockhoff, and R.~H.~B. Christensen, ``{lmerTest} package: Tests in linear mixed effects models,'' \emph{Journal of Statistical Software}, vol.~82, no.~13, pp. 1--26, 2017.

\bibitem{Lienard_2019}
J.-S. Liénard, ``\BIBforeignlanguage{en}{Quantifying vocal effort from the shape of the one-third octave long-term-average spectrum of speech},'' \emph{\BIBforeignlanguage{en}{J. Acoust. Soc. America}}, vol. 146, no.~4, p. EL369–EL375, Oct. 2019.

\bibitem{Titze_Sundberg_1992}
I.~R. Titze and J.~Sundberg, ``\BIBforeignlanguage{en}{Vocal intensity in speakers and singers},'' \emph{\BIBforeignlanguage{en}{J. Acoust. Soc. America}}, vol.~91, no.~5, 1992.

\bibitem{Rilliard_2018}
A.~Rilliard, C.~d’Alessandro, and M.~Evrard, ``\BIBforeignlanguage{en}{Paradigmatic variation of vowels in expressive speech: Acoustic description and dimensional analysis},'' \emph{\BIBforeignlanguage{en}{J. Acoust. Soc. America}}, vol. 143, no.~1, p. 109–122, Jan. 2018.

\bibitem{Boula_2012}
P.~Boula~de Mareüil, A.~Rilliard, and A.~Allauzen, ``\BIBforeignlanguage{en}{A diachronic study of initial stress and other prosodic features in the french news announcer style: Corpus-based measurements and perceptual experiments},'' \emph{\BIBforeignlanguage{en}{Language and Speech}}, vol.~55, no.~2, 2012.

\end{thebibliography}

\end{document}